\documentstyle[12pt]{article}
\begin{document}
\title{RADIATION AND VORTICITY: THE MISSING LINK }
\author{L. Herrera$^1$\thanks{On leave from UCV, Caracas, Venezuela, e-mail: lherrera@usal.es}\\
{$^1$Departamento de  de F\'{\i}sica Teorica e Historia de la  F\'{\i}sica,}\\
Universidad del Pais Vasco, Bilbao, Spain}
\maketitle

\begin{abstract}
In the context of General Relativity,  radiation,  either gravitational or electromagnetic, is closely associated to vorticity of observers world lines. We stress in this letter that  the factor that relates the two phenomena is a circular flow of  energy (electromagnetic) and/or superenergy  on the planes orthogonal to vorticity vector. We also stress the potential relevance of the abovementioned  relationship in experiments to detect gravitational radiation.
\end{abstract}

{\it To Hermann Bondi and Bill Bonnor: Masters and Friends,...and to Michel Bras for his 
gargouillou de jeunes l\'{e}gumes}.

\section{RADIATION AND VORTICITY IN ASYMPTOTICALLY FLAT SYSTEMS}
Radiation (at classical level) is the physical process by means of which the source of the field ``informs'' about any changes in its structure (this includes of course changes in its state of motion).

Thus the knowledge of the field outside the source, on a given hypersurface (spacelike or null) is not sufficient to forecast, {\it modulo} the field equations, the future of the field beyond that hypersurface. This is particularly well illustrated in the Bondi formalism \cite{bondi, 17, 18} (see also \cite{janis} for a  similar discussion in the context of the spin coefficient formalism). It should be stressed  that both formalisms explicitly demand asymptotic flatness.

The information required to forecast the evolution of the system (besides the ``initial'' data) is identified  with radiation itself. In the context of the  Bondi formalism this information is represented by the so called ``news function''. In
other words, whatever happens at the source, leading to changes in the field, it can only
do so by affecting the news function and vice versa. In light of this comment the relationship between the news function and the occurrence of radiation becomes clear. This scheme applies to Maxwell systems in Minkowski spacetime \cite{janis} as well as to Einstein--Maxwell systems \cite{18}.

To reinforce this picture, it turns out that the Bondi mass of a system is constant if and only if there are no news.

On the other hand the vorticity vector ${\bf \Omega}$ associated with the vorticity  of a congruence 
describes the rate of rotation (the proper angular velocity) with respect to the proper time at any point at rest in the rotating frame, relative to the local compass of inertia. Therefore $-{\bf \Omega}$ decribes the rotation of the compass of inertia (gyroscope) with respect to reference particles \cite{6}. 

The intriguing fact is that these two (very different) phenomena are related.

The first piece of this puzzle appeared when it was  established  that gravitational radiation produces vorticity in the congruence of observers, with respect to the compass of inertia    \cite{1}--\cite{5}, implying that a frame dragging effect is associated with gravitational radiation. This result was obtained by proving that in an expansion of inverse powers of  $r$ (where $r$ denotes the null Bondi coordinate), the coefficient of the vorticity  at order  ${\displaystyle \frac 1{r}}$ will vanish if and only if there are no news.

How does the link between vorticity and radiation appears?

 The rationale to link  vorticity and the super-Poynting vector comes from an idea put forward by  Bonnor in order to explain the appearance of vorticity in the spacetime generated by a charged magnetic dipole \cite{8}.  Bonnor observes that for such a system there exists  a non--vanishing component of the Poynting vector, describing a flow of electromagnetic energy round in circles \cite{9}. He then suggests that such a circular flow of energy affects inertial frames by producing vorticity of congruences of particles, relative to the compass of inertia. Later, this conjecture was shown to be valid for a general axially symmetric stationary electrovacuum metric \cite{10}. It is worth mentioning that this ``circular'' flow of electromagnetic energy is absolutely necessary  in order to preserve conservation of angular momentum, as illustrated by Feynman in the ``paradox'' of the rotating disk with charges and a solenoid \cite{9}.

The idea that a similar mechanism might be at the origin  of vorticity in the gravitational case, i.e.  a circular flow of gravitational energy would produce  vorticity, was  suggested for the first time  in \cite{4}. However the nonexistence of   a local and invariant  definition of  gravitational energy, rose at that time the question about what expression for the ``gravitational'' Poynting vector  should be used. Following a suggestion by Roy Maartens we tried in \cite{5} with the super--Poynting vector based on the Bel--Robinson tensor \cite{11}--\cite{14}.

Indeed, as  is known, in classical field theory, energy is a quantity
defined in terms of potentials and their first derivatives. In
General Relativity however, it is impossible to construct a tensor
expressed only through the metric and their first derivatives (in accordance with  the
equivalence principle). Therefore, a local description of
gravitational energy in terms of true invariants (tensors of any
rank) is  not possible within the context of the theory.

Thus, one is left with the following three alternatives:
\begin{itemize}
\item  Looking for a non--local  definition of energy.
\item  Finding a definition based on  pseudo--tensors.
\item  Resorting to a succedaneous definition, e.g.:  superenergy.
\end{itemize}

One example of the last of the above alternatives is superenergy,   which may be defined
from the Bel  or the  Bel--Robinson tensor 
(they both coincide in vacuum), and has been shown to be very useful
when it comes to explaining  a number of  phenomena in the context of
general relativity.

In \cite{5}, we were able to establish the link between  gravitational radiation and vorticity, invoking   a mechanism similar to that proposed by Bonnor for the charged magnetic dipole. Indeed, it was shown  that the resulting vorticity is always associated to a circular flow of superenergy on the plane orthogonal to the vorticity vector. It was later shown that the vorticity appearing in stationary vacuum spacetimes also depends on  the existence of a flow of superenergy on the plane orthogonal to the vorticity vector \cite{7}. Furthermore in \cite{16} it was shown that not only gravitational but also electromagnetic radiation produces vorticity. In this latter case we were able to isolate contributions from both the electromagnetic Poynting vector as well as from the super--Poynting vector.

An important feature to underline is that at order (${\displaystyle \frac 1{r^2}}$) there are contributions to the vorticity with a time dependent term not involving news. This last term represents the class of non-radiative motions discussed by Bondi \cite{bondi} and may be thought to correspond to the tail of the wave, appearing after the radiation process \cite{BoNP}. The obtained expression allows for ``measuring'' (in a gedanken experiment, at least) the wave-tail field. This in turn implies that observing the gyroscope, for a period of time from an initial static situation until after the vanishing of the news, should allow for an unambiguous identification of a gravitational radiation process.

\section{RADIATION AND VORTICITY IN CYLINDRICALLY SYMMETRIC SYSTEMS}
An interesting example related to the kind of problem we are examining here is represented by cylindrically symmetric vacuum non--static spacetimes (Einstein--Rosen). In this latter case asymptotic flatness fails to be fulfilled and Bondi formalism cannot be applied.

Einstein--Rosen (E--R) spacetime \cite{ER} has attracted the attention of researchers for many years (see\cite{Thorne}--\cite{Carmeli} and references therein). Its interest may be understood by recalling that E--R represents the simplest example of a spacetime describing outgoing gravitational waves. 

The line element in this case reads
\begin{equation}ds^2=-e^{2\gamma-2\psi}(dt^2-dr^2)+e^{2\psi}dz^2+r^2 e^{-2\psi}d\phi^2\label{1}\end{equation}
where $\psi=\psi(t,r)$ and $\gamma=\gamma(t,r)$ satisfy the Einstein equations:
\begin{equation}\psi_{tt}-\psi_{rr}-\frac{\psi_r}{r}=0\label{2w}\end{equation}
\begin{equation}\gamma_{t}=2r\psi_{r}\psi_t\label{3}\end{equation}
\begin{equation}\gamma_{r}=r(\psi_{r}^2+\psi_t^2)\label{4}\end{equation}
where indexes stand for differentiation with respect to $t$ and $r$ and $x^0=t, x^1=r, x^2=z, x^3=\phi$.

Now, it is obvious that for  a congruence of observers at rest in the frame of (\ref{1}), the vorticity  vanishes.  From the comments above about the link between radiation and vorticity, this result might  shed some doubts on the radiating character of E--R spacetime. However this is not so.

Indeed, for the super--Poynting vector  we obtain (see \cite{HE} for details)

\begin{eqnarray}P_1&=& e^{3(\psi-\gamma)}[\psi_{rr}(-6\psi_r \psi_t+2r\psi^3_t+6r\psi_t\psi^2_r)-2\psi_{rr}\psi_{tr} \nonumber \\&+&\psi_{tr}(-3\psi^2_t-3\psi^2_r-\frac{\psi_r}{r}+6r\psi_r\psi^2_t+2r\psi^3_r)-8\psi^3_t\psi_r \nonumber\\&+& 3r\psi^5_t+30r\psi^3_t\psi^2_{r}-6\psi^3_r\psi_t-\frac{3\psi^2_r \psi_t}{r}+15r\psi^4_r\psi_t\nonumber \\&-& 6r^2\psi_r\psi^5_t -20r^2\psi^3_t\psi^3_r -6r^2\psi^5_r\psi_t].\label{p4}\end{eqnarray}
and
 \begin{equation}P_2=P_3=0.
\label{p2}\end{equation}
As a simple example, let us now consider a cylindrical source which is static for a period of time until it starts contracting and emits a sharp pulse of radiation traveling outward from the axis. Then, the function $\psi$ can be written as \cite{Carmeli, Herrera}
\begin{equation}\psi=\frac{1}{2 \pi}\int_{-\infty}^{t-r}\frac{f(t^{\prime})dt^{\prime}}{[(t-t^\prime)^2-r^2]^{1/2}} + \psi_{LC},\label{pulse1}\end{equation}
In (\ref{pulse1}) $\psi_{LC}$ represents the Levi-Civita static solution defined by 
\begin{equation}\psi_{LC} = \alpha - \beta \ln{r}, \quad \alpha,\beta \; {\rm constants,}\label{psist}\end{equation}\begin{equation}\gamma_{LC} =\beta^2 \ln{r}\label{gamst}\end{equation}
and  $f(t)$ is a function of time representing the strength of the source of the wave and it is assumed to be of the form
\begin{equation}f(t)=f_0\delta (t),\label{pulse2}\end{equation}
where $f_0$ is a constant and $\delta (t)$ is the Dirac delta function. It can be shown that (\ref{pulse1}) satisfies the wave equation (\ref{2w}).Then we get
\begin{eqnarray}\psi=\psi_{LC}, \;\; t<r; \label{pulse3} \\\psi=\frac{f_0}{2\pi(t^2-r^2)^{1/2}} + \psi_{LC}, \;\; t>r. \label{pulse4}\end{eqnarray}
The function $\psi$,  as well as its derivatives, are regular everywhere except at the wave front determined by the surface $t=r$, followed by a tail decreasing with $t$. 

For the pulse--type solution, using  (\ref{pulse4}), in the limit $t \approx r (t>r)$ (just behind the pulse) we obtain\begin{equation}P^1 \approx \frac{e^{3(\psi-\gamma)}f_0^5 t^5}{2 \pi^5 (t^2-r^2)^{15/2}}\left[\frac{f_0t^2}{\pi(t^2-r^2)^{3/2}}-\frac{7 \beta}{2}\right]\label{sptr}\end{equation}which is positive if\begin{equation}\beta<\frac{2f_0 r^2}{7 \pi (t^2-r^2)^{3/2}}.\label{c2}\end{equation}

Thus, there is a radial flux of super--energy, implying the emission of gravitational waves. On the other hand the absence of vorticity is explained by the fact that the only nonvanishing component of the super--Poynting vector is the radial one. In  other words there is no ``circular'' flow of superenergy in any plane of the $3$-space. 

In the example above the  ``delta'' function behaviour  of first derivatives of $\psi$, related to the choice in (\ref{pulse2}), might shed some doubts about the validity of  such example. In this respect, three comments are in order:
\begin{itemize}
\item It should be emphasized that function  $\psi$ as defined  by (\ref{pulse4}) is regular and well behaved in the whole region $t>r$. Accordingly, it is reasonable to assume that  the definition of a radial component of the super--Poynting vector in such a region exists. 
\item It should be observed that behind the pulse (but not close to it) the radial component of the super--Poynting vector becomes negative. Such a ``strange'' result becomes intelligible when we recall that the field behind the pulse becomes asymptotically (in time), the  static Levi--Civita solution  (\ref{psist}). Therefore, it should be clear that in order to restore the static situation with the same initial $\beta$ the system should absorb some energy, in order to compensate for the energy that has been radiated away (remember   that $\beta$ is related to the mass per unit of length). This is a reminiscense of  the situation that one finds in  classical electrodynamics, for the field of a current along an infinite wire,  in which case,  the Poynting vector describes a flux of electromagnetic energy directed radially into the wire \cite{9}. In that example Feymann dismisses the significance of such an effect by arguing that it is deprived of any physical relevance. However in our case, the presence of a radial inwardly directed flux, is necessary in order to restore the energy  carried away by the pulse and at the same time, to explain the time evolution pattern of the angular velocity of a test particle in a circular motion around the source (see \cite{HE} for a discussion on this point).
\item Finally, beyond the discussion above, it should be stressed that there are only two important facts to retain, namely: on the one hand, the solution describes a pulse of  gravitational waves (it is a solution of (\ref{2w})) moving radially outward (independently on whether or not a corresponding super--Poynting vector can be properly defined), and on the other  hand, and this is the fundamental fact, there are no components of the super--Poynting vector, on the plane orthogonal to the propagation  direction.
\end{itemize}
\section{Conclusions}
We may summarize the main issues addressed in this letter in the following  points:
\begin{itemize}
\item  Gravitational and electromagnetic radiation produce vorticity in asymptotically flat spacetimes, in which  case there is always a circular flow of electromagnetic energy and/or superenergy on at least one plane  of the $3$-space.
\item For unbounded configurations  (e.g. cylindrically symmetric sytems), radiation  does not produce vorticity if there are no  a circular flow of superenergy on any  plane of the $3$-space (e.g. E--R spacetime).
\item The two points above imply that the generation of vorticity by means of radiation is achieved   {\it modulo}  a Bonnor--like mechanism.
\item Such a vorticity should be detected by a gyroscope attached to the world line observer.
\item The potential observational consequences of the generation of vorticity by radiation should be seriously considered. Indeed, the   direct experimental evidence of the existence of the Lense--Thirring effect \cite{22, ciu1, ciu2} brings out the high degree of development achieved in  the required  technology. In the same direction point recent proposals to detect frame dragging by means of ring lasers \cite{rl1, rl1bis, rl2, rl3, rl4}. Also it is worth mentioning the possible use of atom interferometers \cite{at1}, \cite{at2}, \cite{at3}, atom lasers  \cite{atl} and  anomalous spin--precession experiments \cite{sp} to measure vorticity.
\item At the order ${\displaystyle \frac1{r^2}}$  there are contributions to the vorticity from terms  not involving news. Thus let us suppose that at some initial time (retarded) $u = u_0$ the system starts to radiate until $u = u_f$, when the news vanish again. For $u>u_f$ the system is not radiating but  there is a vorticity term of order ${\displaystyle \frac1{r^2}}$  describing the effect of the tail of the wave on the gyroscope. This in turn provides ``observational'' evidence for the violation of the Huygens'sprinciple, a problem largely discussed in the literature (see for example\cite{bondi},\cite{BoNP},\cite{tail} and references therein). It is also important to note that this ``tail'' effect would generate a reception signal  with a very  distinct profile.
\end{itemize}

\end{document}